\font\bbf=cmbx12

\font\srm=cmr9
\font\sit=cmti9
\font\sbf=cmbx9

{\hfill \srm $\copyright$ 1999 The American Institute of Physics}
\vskip 1truein
\centerline{\bbf PROGRESS IN ESTABLISHING A CONNECTION BETWEEN THE}
\centerline{\bbf ELECTROMAGNETIC ZERO-POINT FIELD AND INERTIA}
\bigskip
\centerline{\bbf Bernhard Haisch$^1$ and Alfonso Rueda$^2$}
\bigskip
\centerline{\srm $^1$Solar \& Astrophysics Laboratory, Lockheed Martin, H1-12, B252, 3251
Hanover St., Palo Alto, CA 94304}
\centerline{\srm haisch@starspot.com}
\centerline{\srm $^2$Dept. of  Electrical Eng. and Dept. of Physics \& Astronomy,
California State Univ., Long
Beach, CA 90840}
\centerline{\srm  arueda@csulb.edu}
\bigskip
\centerline{\sit Presented at Space Technology and Applications International Forum
(STAIF-99)}
\centerline{\sit January 31--February 4, 1999, Albuquerque, NM}
\bigskip

{\parindent 0.4truein \narrower \noindent
{\sbf Abstract.}
We report on the progress of a NASA-funded study being carried out at the Lockheed Martin
Advanced Technology Center in Palo Alto and the California State University in Long Beach
to investigate the proposed link between the zero-point field of the quantum vacuum and
inertia.  It is well known that an accelerating observer will experience a bath of
radiation resulting from the quantum vacuum which mimics that of a heat bath, the
so-called Davies-Unruh effect. We have further analyzed this problem of an accelerated
object moving through the vacuum and have shown that the zero-point field will
yield a non-zero Poynting vector to an accelerating observer. Scattering of this radiation
by the quarks and electrons constituting matter would result in an acceleration-dependent
reaction force that would appear to be the origin of inertia of matter (Rueda and Haisch
1998a, 1998b). In the subrelativistic case this inertia reaction force is exactly
newtonian and in the relativistic case it exactly reproduces the well known relativistic
extension of Newton's Law. This analysis demonstrates then that both the ordinary,
$\vec{F}=m\vec{a}$, and the relativistic forms of Newton's equation of motion may be
derived from Maxwell's equations as applied to the electromagnetic zero-point field. We
expect to be able to extend this analysis in the future to more general versions of the
quantum vacuum than just the electromagnetic one discussed herein.

}

\bigskip\centerline{\bbf BACKGROUND}
\medskip\noindent
In July 1998 the Advanced Concepts Office at JPL and the NASA Office of Space Science
sponsored a four-day workshop at Caltech on ``Robotic Interstellar Exploration in the Next
Century.'' The objective was to bring together scientists and
engineers to survey the landscape of possible ideas that could lead to unmanned missions
beyond the Solar System beginning within a timeframe of 40 years. Missions to Kuiper Belt
objects ($>40$ AU), the local interstellar medium beyond the heliopause ($>150$ AU), the
Oort Cloud ($10000-50000$ AU), and ultimately the nearest star system ($\alpha$ Centauri at
270000 AU) were considered.

\medskip\noindent
Present rocket technology falls short of the necessary propulsion requirements by
orders of magnitude. Radically new capabilities are needed, and one approach is by
extreme extrapolation of known technologies.
Ideas presented at the workhop thus included: anti-matter initiated fusion; particle beams
which could be accurately directed to a target vehicle at light-year distances owing to
nanotechology navigation capabilities built into the particles themselves;
current-carrying tether arrays 1000 by 1000 km in size that could generate Lorentz forces
by interaction with the interstellar magnetic field; laser-pushed lightsails that could be
accurately pointed and maintain collimation at stellar distances, etc.

\medskip\noindent
The extreme sizes of structures and the extreme tolerances required for such concepts is
worrisome. Taking the laser-driven sail as an example, let us assume that a mission
propelled by a 1 km diameter lightsail is halfway ($2
\times 10^{13}$ km) to
$\alpha$ Centauri when a beam problem reaches the vehicle. A one part in $10^{13}$
misalignment of the laser {\it which occurred 2 years previously back on earth} is now
reaching the vehicle causing the beam to miss the light sail. Owing to the speed-of-light
limitation, it will be another 2 years before any news of this transmitted by the
spacecraft can reach the earth. It will be yet another 2 years before the correction
from earth will reach the spacecraft. But by then the vehicle may have drifted out of its
trajectory sufficiently owing to the secular effects of interaction with the
interstellar medium (or other causes) that it is still out of the beam. Indeed, any drift
of the vehicle from a line-of-sight trajectory will cause the same uncorrectible problem
in the first place since there is no way to know where the vehicle is ``now'' (in the
sense of where the beam is supposed to hit). This illustrates the inherent problem of
speed-of-light caused time delay in any feedback loop; it would be all too easy --- in
fact probably unavoidable --- to have a mission ``lost in space without a paddle'' due to
the slightest error using this propulsion mechanism.

\medskip\noindent
An alternative approach to extreme and perhaps unrealistic technologies is to
consider what kinds of new physics might leapfrog us beyond this: after all, no amount of
money and energy expended on maximizing information transmission via 19th century pony
express couriers could approach the amount and instantaneity of information transmission by
televison or the internet for example. In this case, the capabilities of ``new physics'' in
electrodynamics utterly superseded mechanics.

\medskip\noindent
Obviously one would like new physics that will permit faster-than-light $(v>c)$ travel and
provide access to unlimited energy. The $v>c$ hope cannot be encouraged, however, owing to
the fundamental conflicts it causes with respect to relativity and causality.
\footnote{$^a$}
{In a famous speech in 1900, Lord Kelvin extolled the near completeness of physics
(owing in no small measure, naturally, to his own Herculean efforts) with only two dark
clouds on the otherwise clear horizon: the blackbody problem and the failure to detect the
ether. At the moment there is arguably a glimmering of two even smaller clouds on the
horizon with respect to relativity and causality that may ultimately point the way to some
conceivable
$v>c$ physics. With respect to relativity, it would be surprising if the rest frame
defined by the cosmic microwave background did not turn out to be somehow special after
all, perhaps opening the door to non-Lorentzian space-time physics. With respect to
causality, the cloud is even wispier. It is an enigma within an anomaly that there is some
credible evidence for human extrasensory perception and that this perception of
information appears not to be dependent upon time, there apparently being no greater
barrier to accessing future information than present information (Jahn et al. 1997). It
would, of course, be unwise to make any interstellar plans on this basis.} However it does
appear that there may be the equivalent of other new physics lurking within the
electrodynamics of the quantum vacuum.

\medskip\noindent
At this time there are four possibilities relevant to future propulsion
technology that have a sufficiently well-developed basis in quantum vacuum physics so as to
warrant further theoretical investigation: extraction of energy, generation of force, and
manipulation of inertia and possibly even of gravitation. {\it The realization of any one
of these would leapfrog beyond all other concepts of interstellar travel presented at the
workshop, and this was the topic of an invited presentation at the Caltech conference by
Haisch.}

\bigskip\centerline{\bbf THE ZERO-POINT FIELD OF THE QUANTUM VACUUM}
\medskip\noindent
A NASA-funded research effort has been underway since 1996 at the Lockheed Martin Advanced
Technology Center in Palo Alto and at the California State University in Long Beach to
explore the physics of the quantum vacuum and its possible long-term potential
applications. That effort is a follow-on to previous work suggesting a relationship between
the electromagnetic zero-point field of the quantum vacuum and inertia (Haisch, Rueda
and Puthoff, 1994).

\medskip\noindent
In the conventional interpretation of quantum theory, an electromagnetic zero-point field
arises as a consequence of the Heisenberg uncertainty relation as applied to each mode of
the electromagnetic field. No oscillator can ever be
brought completely to rest due to quantum fluctuations. The minimum energy for a mechanical
oscillator whose natural frequency is $\nu$ is $E=h\nu/2$. Each mode of the
electromagnetic field also acts as an oscillator. Thus for any frequency, $\nu$, direction,
$\vec{k}$, and polarization state, $\sigma$, there is a minimum energy of $E=h\nu/2$ in the
electromagnetic field. Summing up all of these modes, each with its $E=h\nu/2$ of energy,
results in an electromagnetic ground state of energy that should permeate the entire
universe: the electromagnetic zero-point field, or ZPF. (The term ZPF refers to either the
zero-point fields or equivalently zero-point fluctuations of the electromagnetic quantum
vacuum; the term ZPE refers to the energy content of the electromagnetic quantum vacuum.)
All other natural or artificial electromagnetic radiation would sit on top of this very
energetic ground state.

\medskip\noindent
The volumetric density of modes
between frequencies $\nu$ and
$\nu+d\nu$ is given by the density of states function
$N_{\nu}d\nu=(8\pi\nu^2/c^3)d\nu$. Using this density of states function and the minimum
energy, $h\nu/2$, that we call the zero-point energy per state one can calculate the
ZPF spectral energy density:
$$\rho(\nu)d\nu={8\pi\nu^2 \over c^3} {h\nu
\over 2} d\nu .
\eqno(1)
$$

\medskip\noindent
It is instructive to write the expression for zero-point spectral energy
density side by
side with blackbody radiation:

$$\rho(\nu,T)d\nu={8\pi\nu^2 \over c^3} \left( {h\nu \over e^{h\nu /kT} -1}
+{h\nu
\over 2}\right) d\nu .
\eqno(2)
$$
\medskip\noindent
The first term (outside the parentheses) represents the mode density, and the
terms inside the parentheses are the average energy per mode of thermal
radiation at temperature $T$ plus the zero-point energy, $h\nu/2$, which has no
temperature dependence. Take away all thermal energy by formally letting $T$
go to zero, and one is still left with the zero-point term: the ground state of the
electromagnetic quantum vacuum. The laws of quantum mechanics as applied to electromagnetic
radiation force the existence of a background sea of zero-point-field (ZPF) radiation.

\medskip\noindent
Zero-point radiation is taken to result from quantum laws. It is
traditionally assumed in quantum theory, though, that the ZPF can for practical
purposes be
ignored or subtracted away. The foundation of the discipline in physics known as
stochastic electrodynamics (SED) is the exact opposite (see e.g. de la Pe\~na and Cetto
1996 for a thorough review of SED). It is assumed in SED that the ZPF is as real as any
other electromagnetic field. As to its origin, the
assumption is that zero-point radiation simply came with
the Universe.
The justification for this is that if one assumes that all of space is
filled with ZPF radiation, a
number of quantum phenomena may be explained purely on the basis of
classical physics
including the presence of background electromagnetic fluctuations provided
by the ZPF. The
Heisenberg uncertainty relation, in this view, becomes then not a result of
the existence
of quantum laws, but of the fact that there is a universal perturbing ZPF
acting on everything. The original motivation for developing SED was to see
whether the
need for quantum laws separate from classical physics could thus be
obviated entirely.

\medskip\noindent
Philosophically, a universe filled --- for reasons unknown --- with a ZPF
but with only one
set of physical laws (classical physics consisting of mechanics and
electrodynamics), would
appear to be on an equal footing with a universe governed --- for reasons
unknown ---  by
two distinct physical laws (classical and quantum). In terms of physics,
though, SED
and quantum electrodynamics, QED, are not on an equal footing, since SED has been
successful in providing a satisfactory alternative to only some quantum phenomena
(although this success does
include a classical ZPF-based derivation of the all-important blackbody
spectrum, cf.
Boyer 1984). Some of this is simply due to lack of effort: The ratio of
man-years devoted
to development of QED is several orders of magnitude greater than the expenditure
so far on SED.

\medskip\noindent
There is disagreement about whether this zero-point field should be regarded as real or
virtual. A number of well-established phenomena such as the Casimir force and the Lamb
shift are equally well explained in terms of either the action of a real ZPF or simply the
quantum fluctuations of particles. This paradox is discussed in some detail by Milonni
(1988). It is clearly essential to determine how ``real'' the zero-point field is.

\bigskip\centerline{\bbf ENERGY EXTRACTION FROM THE QUANTUM VACUUM}

\medskip\noindent
In the
early part of this century the discovery of radioactivity seemed to violate the law of
energy conservation. Heat and radiation appeared to be continuously given off as if by a
source of ``free energy'' in certain elements (e.g. radium, uranium). The resolution came
with the understanding that mass was being converted into energy via the
$E=mc^2$ relationship of special relativity in a process of spontaneous decay of unstable
elements, the naturally occuring radioisotopes. The low-level energy emitted by decay of
natural or artificially-created radioisotopes is of limited use. However with the
successful demonstration of fission in the 1940's, nuclear engineering became possible
allowing us to tap a much more powerful mode of atomic energy release .

\medskip\noindent
Chemical energy production, as in the burning of petroleum-based fuels or conventional
rocket propulsion, taps the energy in the orbital electrons of atoms. Atomic energy
generation taps the binding energy of the nuclear constituents of atoms. Is it possible to
tap yet a deeper (and potentially much more powerful) source of energy: the ZPE of the
electromagnetic quantum vacuum? There are two issues that bear on this.

\medskip\noindent
It is often assumed that attempting to tap the energy of the vacuum must violate
thermodynamics. One cannot extract thermal energy from a reservoir at temperature,
$T_1$, if the environment is at temperature, $T_2>T_1$. While it is
true that the ZPE is the energy remaining when all energy sources have been removed and the
temperature reduced to $T=0$ K, the ZPE is not a thermal reservoir. It has very different
characteristics than ordinary heat. The thermodynamics of energy extraction from the
quantum vacuum have been analyzed by Cole and Puthoff (1993). They conclude as follows:

\medskip
{\parindent 0.4truein \narrower \noindent
Relatively recent proposals have been made in the literature for extracting energy and
heat from electromagnetic zero-point radiation via the use of the Casimir force. The basic
thermodynamics involved in these proposals is analyzed and clarified here, with the
conclusion that, yes, in principle, these proposals are correct. Technological
considerations for the actual application and use are not examined here, however.

}

\medskip\noindent
{\it If} the zero-point field is a real electromagnetic ground state, then there is no
inconsistency with present-day physics in the possibility of tapping this energy. Indeed,
it has been proposed that certain astrophysical processes are driven by the natural
extraction of such energy (cf. Rueda, Haisch and Cole 1995 and references therein) and an
ideal experiment has been proposed by Forward (1984) that clearly demonstrates the
conceptual possibility of extracting vacuum energy.

\medskip\noindent
A second point to consider is that the orbital energy of electrons and the $E=mc^2$
relationship itself may both ultimately be traceable to zero-point energy. While
limited so far to the single, simple case of the ground-state of hydrogen, the work of
Boyer (1975) and Puthoff (1987) suggests that electron energy levels may be stabilized
against radiative collapse by interaction with the quantum vacuum. This would in principle
link chemical energy to the energy of the ZPF.

\medskip\noindent
In his preliminary development of the
Sakharov conjecture on gravity as a ZPF-induced force, Puthoff (1989) suggests that the
$E=mc^2$ relationship reflects the kinetic energy of {\it zitterbewegung} (Schr\"odinger's
term) which originates in the fluctuations induced by the ZPF on charged particles (quarks
and electrons). In other words, instead of expressing a relationship between mass and
energy, the
$E=mc^2$ relationship tells us how much ZPF-driven energy is associated with a given
particle. When this energy is liberated it is thus not really a transformation of mass
into energy, rather a release of zero-point energy associated with this quantum motion
known as {\it zitterbewegung}. This would in principle link atomic
energy to the energy of the ZPF. (An attempt at a classical visualization of this motion
from the SED viewpoint may be found in Rueda, 1993).

\medskip\noindent
{\it If} the above interpretations prove to be valid and it is ultimately the energy of
the ZPF that is being tapped in chemical and atomic energy production processes, then it
is not inconceivable that other channels will be found to liberate energy from the
quantum vacuum. Since the ZPF must be universal, a propulsion drive energized by the ZPF
would have access to unlimited ``fuel'' anywhere\dots the ZPE thus providing the ultimate
energy source.

\bigskip\centerline{\bbf IN-SITU GENERATION OF FORCES}

\medskip\noindent
The existence of the Casimir force --- an attraction between uncharged conducting plates
--- is now well established. Measurements by Lamoreaux (1997) are in agreement with
theoretical predictions to within a few percent. A particularly simple interpretation
involving the ZPF was presented by Milonni, Cook and Goggin (1988):

\medskip
{\parindent 0.4truein \narrower \noindent
We calculate the vacuum-field radiation pressure on two parallel, perfectly conducting
plates. The modes outside the plates push the plates together, those confined between the
plates push them apart, and the net effect is the well-known Casimir force.

}

\medskip\noindent
In other words, the electromagnetic boundary conditions on the two plates exclude a
certain amount of ZPF from the cavity in between. This results in an overpressure from the
ZPF outside, which then acts as a pressure on the plates.

\medskip\noindent
This is one interpretation: call it the ZPF radiation pressure model. It is also possible
to interpret the force as ``a macroscopic manifestation of the retarded van der Waals force
between two neutral polarizable particles'' (Milonni, Cook and Goggin 1988), i.e. a
quantum effect involving the particles in the plates (see Milonni 1982).

\medskip\noindent
The speculation concerning propulsion is that the ZPF radiation pressure model of the
Casimir force is physically correct and that it may be possible to construct some wall or
cavity that interacts with the ZPF differently on one side than on the other. If that were
possible, one would have in effect a ZPF-sail that could provide a propulsive force
anywhere in space.

\bigskip\centerline{\bbf MODIFICATION OF INERTIAL AND GRAVITATIONAL MASS}

\medskip\noindent
The ultimate capability enabling practical interstellar travel would be the ability to tap
the energy of the vacuum while at the same time modifying the inertia of the spacecraft.
Reducing inertia of a spacecraft would allow higher velocity for the same expenditure of
energy and more rapid acceleration without damage to the structure owing to the reduction
of inertial forces. The absolute limit would be acceleration to velocity $c-\epsilon$ in
time
$\delta$ where both $\epsilon$ and $\delta$ approach zero.

\medskip\noindent
Until recently there was absolutely no basis in physics for even considering such a
possibility. While such a possibility still appears to be remote, there now does exist a
basis in physics to at least begin to explore this concept.

\medskip\noindent
In 1994 Haisch, Rueda and Puthoff published a paper, ``Inertia as a zero-point
field Lorentz Force,'' in which a substantive mathematical analysis indicated that the
inertia of matter could be interpreted as an electromagnetic reaction force originating
in the quantum vacuum. This concept has now been redeveloped by Rueda and Haisch (1998a,
1998b) in a way that is both mathematically simpler while at the same time yielding a
properly covariant relativistic result. This is encouraging that we are on the right track.

\medskip\noindent
In a frame at rest or in uniform motion, the ZPF is uniform and isotropic. This is due to
the Lorentz invariance of the ZPF spectrum. (The spectral cutoff of the spectrum would not
be Lorentz invariant, but if the inertia interaction takes place at a frequency or
resonance far from the cutoff, this would not matter.) However in an accelerated frame the
ZPF becomes asymmetric. Rueda and Haisch have shown that the Poynting vector --- which
characterizes the radiative energy flux  --- becomes non-zero in an accelerating frame. If
the quarks and electrons in matter undergoing acceleration scatter this ZPF flux, a
reaction force will arise that is proportional to accleration. This is proposed to be the
origin of inertia of matter. {\it Inertia is not an innate property of matter; it is an
acceleration-dependent electromagnetic reaction force.}

\medskip\noindent
The principle of equivalence mandates that gravitational and inertial mass must be the
same. Therefore, if inertial mass is electromagnetic in origin, then gravitational mass
must also be electromagnetic in some fashion. A preliminary development of a gravitational
analysis based on the electrodynamics of the ZPF has been made by Puthoff (1989).
Subsequent critiques have pointed out some deficiencies in this analysis. Nevertheless, it
is encouraging that the ZPF-related parameters that determine ``mass'' turn out to be
identical in the inertia and gravitation analyses (see Appendix A in Haisch, Rueda and
Puthoff 1994). From this view, the ZPF acts as a mediator of a gravitational force, but
cannot itself gravitate, hence would not result in an unacceptably large cosmological
constant (Haisch and Rueda 1997).

\medskip\noindent
We now have a theoretical basis to explore the possibility that electrodynamics may be
used to modify the quantum vacuum in some way so as to alter inertia and/or gravitation.
It would be prudent to continue such investigations.
\vfill\eject
\bigskip\centerline{\bbf ACKNOWLEDGMENTS}
\medskip\noindent
We acknowledge support of NASA contract NASW-5050 for this work. BH also
acknowledges the
hospitality of Prof. J. Tr\"umper and the Max-Planck-Institut where some of
these ideas
originated during several extended stays as a Visiting Fellow. AR
acknowledges
stimulating discussions with Dr. D. C. Cole.

\bigskip\centerline{\bbf REFERENCES}
\medskip
{

\bigskip
\parskip=0pt plus 2pt minus 1pt\leftskip=0.25in\parindent=-.25in

Boyer, T. H., ``Random electrodynamics: The theory of classical electrodynamics with
classical electromagnetic zero-point radiation,'' {\it Phys. Rev. D}, {\bf 11}, 790 (1975).

Boyer, T. H., ``Derivation of the blackbody radiation spectrum from the equivalence
principle in classical physics with classical electromagnetic zero-point radiation,'' {\it
Phys. Rev. D}, {\bf 29}, 1096 (1984).

Cole, D.C. and Puthoff, H.E., ``Extracting energy and heat from the vacuum,'' {\it Phys.
Rev. E}, {\bf 48}, 1562 (1993).

de la Pe\~na, L. and Cetto, A.M. {\it The Quantum Dice: An Introduction to Stochastic
Electrodynamics}, Kluwer Acad. Publishers, Dordrecht, (1996).

Forward, R., ``Extracting electrical energy from the vacuum by cohesion of charged
foliated conductors,'' {\it Phys. Rev. B}, {\bf 30}, 1700 (1984).

Haisch, B. and Rueda, A., ``Reply to Michel's `Comment on Zero-Point Fluctuations and the
Cosmological Constant','' {\it Astrophys. J.}, {\bf 488}, 563 (1997).

Haisch, B., Rueda, A. and Puthoff, H.E. (HRP),  ``Inertia as a zero-point-field Lorentz
force,'' {\it Phys. Rev. A}, {\bf 49}, 678 (1994).

Jahn, R.G, Dunne, B.J., Nelson, R.D., Dobyns, Y.H. and Bradish, G.J., ``Correlations of Random Binary Sequences With Pre-stated Operator
Intention: A Review of a 12-year Program,'' {\it J. Scientific
Exploration}, {\bf 11}, 345 (1997).

Lamoreaux, S.K., ``Demonstration of the Casimir Force in the 0.6 to 6 $\mu$m Range,'' {\it
Phys. Rev. Letters}, {\bf 78}, 5 (1997)

Milonni, P.W., ``Casimir forces without the vacuum radiation field,'' {\it Phys. Rev. A},
{\bf 25}, 1315 (1982).

Milonni, P.W., ``Different Ways of Looking at the Electromagnetic Vacuum,'' {\it Physica
Scripta}, {\bf T21}, 102 (1988).

Milonni, P.W., Cook, R.J. and Goggin, M.E.,  ``Radiation pressure from the vacuum:
Physical interpretation of the Casimir force,'' {\it Phys. Rev. A}, {\bf 38}, 1621 (1988).

Puthoff, H.E.,  ``Ground state of hydrogen as a zero-point-fluctuation-determined
state,'' {\it Phys. Rev. D}, {\bf 35}, 3266 (1987).

Puthoff, H.E., ``Gravity as a zero-point-fluctuation force,'' {\it Phys. Rev. A}, {\bf 39},
2333 (1989).

Rueda, A.,  ``Stochastic Electrodynamics with Particle Structure: Part I  -- Zero-point
induced Brownian Behaviour,'' {\it Found. Phys. Letters}, {\bf 6}, 75 (1993); and
``Stochastic Electrodynamics with Particle Structure: Parts II -- Towards a Zero-point
induced Wave Behaviour,'' {\bf 6}, 193 (1993).

Rueda, A. and Haisch, B., ``Inertia as reaction of the vacuum to accelerated motion,'' {\it
Physics Letters A}, {\bf 240}, 115 (1998a).

Rueda, A. and Haisch, B., ``Contribution to inertial mass by reaction of the vacuum to
accelerated motion,'' {\it Foundations of Physics}, {\bf 28}, 1057 (1998b).

Rueda, A., Haisch, B. and Cole, D. C., ``Vacuum Zero-Point Field Pressure Instability
in Astrophysical Plasmas and the Formation of Cosmic Voids,'' {\it Astrophys. J.}, {\bf
445}, 7 (1995).

}

\bye